\documentstyle[12pt]{article}

\begin{document}
{\sf \begin{center} \noindent
{\Large \bf Magnetic dynamo C-flows in Riemannian compact manifold as generalized Arnold's metric}\\[3mm]

by \\[1cm]

{\sl L.C. Garcia de Andrade}\\

\vspace{0.5cm} Departamento de F\'{\i}sica
Te\'orica -- IF -- Universidade do Estado do Rio de Janeiro-UERJ\\[-3mm]
Rua S\~ao Francisco Xavier, 524\\[-3mm]
Cep 20550-003, Maracan\~a, Rio de Janeiro, RJ, Brasil\\[-3mm]
Electronic mail address: garcia@dft.if.uerj.br\\[-3mm]
\vspace{2cm} {\bf Abstract}
\end{center}
\paragraph*{}
It is shown that C-flows in Riemannian three-dimensional compact
manifold can be naturally considered as generalized dynamo Arnold's
Riemann metric in compact manifolds, the so-called cat map dynamo.
The generalized solution of self-induction equation in the
background of this metric shows that one is allowed to consider
stretching along both directions of the flow, instead of compressed
in one direction and stretched in the other such as in Arnold's
dynamo. Though this solution can be considered as unrealistic,at
least for incompressible flows, there is another generalized
solution which considers distinct stretch and compression or
exponential damping are anisotropic, and the dynamo flow is
compressed and stretched non-uniformly along distinct directions.
Riemann curvature tensor components are computed by making use of
Cartan's calculus of differential forms.\vspace{0.5cm} \noindent
{\bf PACS numbers:} \hfill\parbox[t]{13.5cm}{02.40.Hw:Riemannian
geometries}
\newpage

\section{Introduction}
 One of the most important issues in the investigation of magnetic structures in solar and plasma physics is the study of
 existence of realistic magnetic dynamos \cite{1}. On this track an important role has been played by less realistic
 dynamo maps in compact Riemannian manifolds such as torus. Among this ,so to speak more naive dynamo examples, the cat
 fast dynamo map invesigated by Arnold et al \cite{2} and later further
 developed by Childress and Gilbert \cite{3}. This dynamo
 flow makes use of a Riemann metric which is stretched along one direction and compressed in the other. More recently
 conformal mappings to Arnold´s metric \cite{4} have shown also to yield new dynamo solutions. The conformal geodesic
 Anosov \cite{5} flows, of negative constant Riemannian curvature, were also investigated by Vishik \cite{6} and Friedlander and Vishik \cite{7}. Using other appoach using the stretching
 of fluid particles was recently investigated by Kambe´s \cite{8} where the sectional curvature does not depend on
 the flow speed but just on the geometrical quantities of the flow. More recently Chicone and Latushkin \cite{9} have
 provided us with an elementary proof of the fact that the geodesic flow on a unit tangent bundle of a two dimensional
 manifold, constant negative curve provides an example of a kinematic fast magnetic dynamo problem. Taking the advantage of
 the Chicone-Latushkin theorem \cite{10},in this paper it is shown that a
 generalized class of Arnold's dynamos can be obtained by the
 auxiliary $C-flows$ ,where C case stands for classical. Though all
 these in dynamos have the common feature of being inspired by the
 stretch-twist and fold Vainshtein-Zeldovich \cite{11} method, and
 generally, physically realistic dynamos possesses stretching in one
 d irection and compression in the other, the generalized $C-flows$
 dynamos considered here may be considered as stretched in both
 coordinates of the compact Riemannian manifold, or either
 compressed in both directions, the present dynamo solution also
 comtemplates distinct stretch and compression in distinct
 direction, which one could call non-uniform stretching \cite{12}.
 Actually is easy to show that Arnold's dynamo metric is a
 particular case of the $C-flow$ metric. The paper is organised as
 follows: Section II presents a review of Arnold dynamo solution. Section III presents the dynamo solution of the
 self-induction equation in compact Riemannian space. In section IV the computation of the Riemann curvature components is
 computed by making use of Cartan's calculus of differential forms and V presents the conclusions.
 \newpage
 \section{Arnold's dynamos in Riemannian manifolds}
 The Arnold metric line element can be
 defined as \cite{2}
 \begin{equation}
 ds^{2}=e^{-2{\lambda}z}dp^{2}+e^{2{\lambda}z}dq^{2}+dz^{2}
 \label{1}
\end{equation}
which describes a dissipative dynamo model on a 3D Riemannian
manifold. By dissipative here, we mean that contrary to the previous
section, the resistivity $\eta$ is small but finite. The flow build
on a toric space in Cartesian coordinates $(p,q,z)$ given by
$T^{2}\times[0,1]$ of the two dimensional torus. The coordinates p
and q are build as the eigenvector directions of the toric cat map
in ${\cal R}^{3}$ which possesses eigenvalues as
${\chi}_{1}=\frac{(3+\sqrt{5})}{2}>1$ and
${\chi}_{2}=\frac{(3-\sqrt{5})}{2}<1$ respectively. This represents
a simple global translation and is not changed at every point in the
manifold. Let us now recall the Arnold et al \cite{4} vector
analysis forms ,definition of a orthogonal basis in the Riemannian
manifold ${\cal{M}}^{3}$
\begin{equation}
\vec{e}_{p}=e^{{\lambda}z}\frac{{\partial}}{{\partial}p}\label{2}
\end{equation}
\begin{equation}
\vec{e}_{q}=e^{-{\lambda}z}\frac{{\partial}}{{\partial}q}\label{3}
\end{equation}
\begin{equation}
\vec{e}_{z}=\frac{{\partial}}{{\partial}q}\label{4}
\end{equation}
Assume a magnetic vector field $\vec{B}$ on M
\begin{equation}
\vec{B}=B_{p}\vec{e}_{p}+B_{q}\vec{e}_{q}+B_{z}\vec{e}_{z}\label{5}
\end{equation}
The vector analysis formulas in this frame are
\begin{equation}
{\nabla}f=[e^{{\lambda}z}{\partial}_{p}f,e^{-{\lambda}z}{\partial}_{q}f,{\partial}_{z}f]\label{6}
\end{equation}
where f is the map function $f:{\cal{R}}^{3}\rightarrow{\cal{R}}$.
The Laplacian is given by
\begin{equation}
{\Delta}f={\nabla}^{2}f=[e^{2{\lambda}z}{{\partial}_{p}}^{2}f+e^{-2{\lambda}z}{{\partial}_{q}}^{2}f+{{\partial}_{z}}^{2}f]
\label{7}
\end{equation}
while the divergence is given by
\begin{equation}
{\nabla}.\vec{B}=div\vec{B}=div[B_{p}\vec{e}_{p}+B_{q}\vec{e}_{q}+B_{z}\vec{e}_{z}]=[e^{{\lambda}z}{{\partial}_{p}}B_{p}+
e^{-{\lambda}z}{{\partial}_{q}}B_{q}+{{\partial}_{z}}B_{z}]\label{8}
\end{equation}
Thus one may write
\begin{equation}
div{\vec{e}}_{p}=div{\vec{e}}_{q}=div{\vec{e}}_{z}=0\label{9}
\end{equation}
and the curl is written as
\begin{equation}
curl\vec{B}=curl[B_{p}\vec{e}_{p}+B_{q}\vec{e}_{q}+B_{z}\vec{e}_{z}]\label{10}
\end{equation}
where
\begin{equation}
curl_{p}\vec{B}=e^{-{\lambda}z}({\partial}_{q}B_{z}-{\partial}_{z}(e^{{\lambda}z}B_{q}))\label{11}
\end{equation}
\begin{equation}
curl_{q}\vec{B}=-e^{{\lambda}z}({\partial}_{p}B_{z}-{\partial}_{z}(e^{-{\lambda}z}B_{p}))\label{12}
\end{equation}
\begin{equation}
curl_{z}\vec{B}=e^{{\lambda}z}{\partial}_{p}B_{q}-e^{-{\lambda}z}{\partial}_{q}B_{p}\label{13}
\end{equation}
and
\begin{equation}
curl{\vec{e}}_{p}=-{\lambda}{\vec{e}}_{q}\label{14}
\end{equation}
\begin{equation}
curl{\vec{e}}_{q}=-{\lambda}{\vec{e}}_{p}\label{15}
\end{equation}
\begin{equation}
curl{\vec{e}}_{z}=0\label{16}
\end{equation}
The Laplacian operators of the frame basis are
\begin{equation}
{\Delta}{\vec{e}}_{p}=-curlcurl{\vec{e}}_{p}=-{\lambda}^{2}{\vec{e}}_{p}
\label{17}
\end{equation}
\begin{equation}
{\Delta}{\vec{e}}_{q}=-curlcurl{\vec{e}}_{q}=-{\lambda}^{2}{\vec{e}}_{q}
\label{18}
\end{equation}
\begin{equation}
{\Delta}{\vec{e}}_{z}=0 \label{19}
\end{equation}
from these expressions Arnold et al \cite{2} were able to build the
self-induced equation in this Riemannian manifold as
\begin{equation}
{\partial}_{t}B_{p}+v{\partial}_{z}B_{p}=-{\lambda}vB_{p}+{\eta}[{\Delta}-{\lambda}^{2}]{B}_{p}-
2{\lambda}e^{{\lambda}z}{\partial}_{p}B_{z} \label{20}
\end{equation}
\begin{equation}
{\partial}_{t}B_{q}+v{\partial}_{z}B_{q}=+{\lambda}vB_{q}+{\eta}[{\Delta}-{\lambda}^{2}]{B}_{p}-
2{\lambda}e^{-{\lambda}z}{\partial}_{q}B_{z} \label{21}
\end{equation}
\begin{equation}
{\partial}_{t}B_{z}+v{\partial}_{z}B_{z}={\eta}[{\Delta}-2{\lambda}{\partial}_{z}]B_{z}
\label{22}
\end{equation}
Decomposing the magnetic field on a Fourier series, Arnold et al
were able to yield the following solution
\begin{equation}
b(p,q,z.t)=e^{{\lambda}vt}b(p,q,z-vt,0) \label{23}
\end{equation}
where $B(x,y,z,t)=b(p,q,z,t)$ and the fast dynamo limit $\eta=0$ was
used. Now with these formulas at hand, we are able to compute the
new solution of the self-induced magnetic equation in the background
of Riemann Arnold's line element, which can be given in the next
section.
\section{Dynamo $C-flows$ as generalized cat map
metric} Earlier Arnold and Avez \cite{13} investigated the auxiliary
$C-flow$ metric given by
\begin{equation}
ds^{2}={{\lambda}_{1}}^{2z}dp^{2}+{{\lambda}_{2}}^{2z}dq^{2}+dz^{2}
\label{24}
\end{equation}
where now coordinates p and q are given by the global
transformations
\begin{equation}
p=[{\lambda}_{1}x+(1-{\lambda}_{1})y] \label{25}
\end{equation}
\begin{equation}
q=[{\lambda}_{2}x+(1-{\lambda}_{2})y] \label{26}
\end{equation}
The inverse transformations are easily obtainded as
\begin{equation}
x=\frac{[(1-{\lambda}_{2})p-(1-{\lambda}_{1})q]}{\beta} \label{27}
\end{equation}
\begin{equation}
y=\frac{[{\lambda}_{1}p-{\lambda}_{2}q]}{\beta} \label{28}
\end{equation}
where
${\beta}:=({\lambda}_{1}+{\lambda}_{2}-2{\lambda}_{1}{\lambda}_{2})$.
Note that when ${\lambda}_{1}:={{\lambda}_{2}}^{-1}$ the $C-flow$
metric under the above coordinate maps reduces to the Arnold's cat
fast dynamo metric. The vector analysis formulas, in the $C-flows$
metric reads
\begin{equation}
{\nabla}=({{\lambda}_{1}}^{-z}{\partial}_{p},{{\lambda}_{2}}^{-z}{\partial}_{q},{\partial}_{z})
\label{29}
\end{equation}
\begin{equation}
{\nabla}.\vec{B}=
{{\lambda}_{1}}^{-z}{\partial}_{p}B_{p}+{{\lambda}_{2}}^{-z}{\partial}_{q}B_{q}+{\partial}_{z}B_{z}
\label{30}
\end{equation}
Here the general expression for the flow is
\begin{equation}
\vec{u}:=v_{p}{{\lambda}_{1}}^{-z}\vec{e}_{p}+v_{q}{{\lambda}_{2}}^{-z}\vec{e}_{q}
+\vec{e}_{z}\label{31}
\end{equation}
To simplify matters one shall adopt the Childress-Gilbert \cite{3}
choice $\vec{u}:=\vec{e}_{z}$. From this choice it is easy to note
that
\begin{equation} (\vec{B}.{\nabla})\vec{e}_{z}=0
\label{32}
\end{equation}
and
\begin{equation} (\vec{u}.{\nabla})\vec{B}={\partial}_{z}[B_{p}{{\lambda}_{1}}^{-z}\vec{e}_{p}+
{{\lambda}_{2}}^{-z}B_{q}+B_{z}\vec{e}_{z}] \label{33}
\end{equation}
Taking the definitions ${\mu}_{A}=log{{\lambda}_{A}}$ where
($A=1,2$), one is able to write the self-induction equation as
\begin{equation}
{\partial}_{t}B_{p}+{\partial}_{z}{B}_{p}-{\mu}_{1}B_{p}={\eta}({\nabla}^{2}-{{\mu}_{1}}^{2})B_{p}-{\mu}_{1}{{\lambda}_{1}}^{-z}
{\partial}_{z}B_{p} \label{34}
\end{equation}
\begin{equation}
{\partial}_{t}B_{q}+{\partial}_{z}{B}_{p}-{\mu}_{2}B_{q}={\eta}({\nabla}^{2}-{{\mu}_{2}}^{2})B_{q}-{\mu}_{2}{{\lambda}_{2}}^{-z}
{\partial}_{z}B_{q} \label{35}
\end{equation}
\begin{equation}
{\partial}_{t}B_{z}+{\partial}_{z}{B}_{z}={\eta}({\nabla}^{2}-2{{\mu}_{1}}^{2})(-{{\lambda}_{1}}^{-z}B_{p}+
{{\lambda}_{2}}^{-z}{\partial}_{q}B_{q}) \label{36}
\end{equation}
Let us consider the case of the highly conductive non-dissipative
flow , where the resistivity $\eta$ vanishes and these equations
reduce to
\begin{equation}
{\partial}_{t}B_{p}+{\partial}_{z}{B}_{p}-{\mu}_{1}B_{p}=0
\label{37}
\end{equation}
\begin{equation}
{\partial}_{t}B_{q}+{\partial}_{z}{B}_{q}-{\mu}_{2}B_{q}=0
\label{38}
\end{equation}
\begin{equation}
{\partial}_{t}B_{z}+{\partial}_{z}{B}_{z}=0 \label{39}
\end{equation}
Thus these equations yield the solution
\begin{equation}
(B_{p},B_{q},B_{z})(p,q,z,t)=({{\lambda}_{1}}^{t}{B_{p}}^{0},{{\lambda}_{2}}^{t}{B_{q}}^{0},B_{z})(p,q,z-t)
\label{40}
\end{equation}
From this expression is easy to see that if one makes the assumption
above about the constants ${\lambda}_{A}$ one obtains the
Childress-Gilbert solution for fast dynamo metric in the limit of
non-dissipative flows. Thus this solution represents the generalized
case where the stretch can be given in both directions
${\lambda}_{1}>0$ and ${\lambda}_{2}>0$ , compressed along both
directions, or ${\lambda}_{1}>0$ and ${\lambda}_{2}<0$, which are
however, unrealistic or even unphysical models, while the last
option, ${\lambda}_{1}>0$ and ${\lambda}_{2}<0$ or vice-versa
represents the stretch and compression along the $(p,q)$ directions
is a physical dynamo solution which is more general than Arnold's
solution since now the stretch is non-uniform.
\newpage
\section{Riemann curvature of $C-flow$ dynamos}
Due to the importance of constant negative curvature of geodesic
Anosov flows for dynamo maps, in this section one computes the
Riemann curvature for C-flows. The C-flow generalization to Arnold
metric can be expressed in terms of the Cartan \cite{14} frame basis
form ${\omega}^{i}$ $(i=1,2,3)$, as
\begin{equation}
ds^{2}=({{\omega}^{p}})^{2}+({{\omega}^{q}})^{2}+({{\omega}^{z}})^{2}\label{41}
\end{equation}
The basis form are write as
\begin{equation}
{\omega}^{p}={{\lambda}_{1}}^{z}dp \label{42}
\end{equation}
\begin{equation}
{\omega}^{q}={{\lambda}_{2}}^{z}dq \label{43}
\end{equation}
and
\begin{equation}
{\omega}^{z}=dz \label{44}
\end{equation}
Applyication of the exterior differentiation of this basis form
yields
\begin{equation}
d{\omega}^{p}= {\omega}_{z}{\wedge}{\omega}_{p} \label{45}
\end{equation}
\begin{equation}
d{\omega}^{z}=0 \label{46}
\end{equation}
by Poincare lemma, and
\begin{equation}
d{\omega}^{q}= {\omega}_{z}{\wedge}{\omega}_{q} \label{47}
\end{equation}
Assuming that our manifold is Riemannian, the Cartan torsion
$2-forms$ of non-Riemannian geometry vanishes, and one obtains, from
Cartan first structure equations
\begin{equation}
T^{p}=0=d{\omega}^{p}+{{\omega}^{p}}_{q}{\wedge}{\omega}^{q}+
{{\omega}^{p}}_{z}{\wedge}{\omega}^{z}\label{48}
\end{equation}
\begin{equation}
T^{q}=0=d{\omega}^{q}+{{\omega}^{q}}_{p}{\wedge}{\omega}^{p}+
{{\omega}^{p}}_{z}{\wedge}{\omega}^{z}\label{49}
\end{equation}
\begin{equation}
T^{z}=0={{\omega}^{p}}_{q}{\wedge}{\omega}^{q}+
{{\omega}^{p}}_{z}{\wedge}{\omega}^{z}\label{50}
\end{equation}
the following constraints
\begin{equation}
{\omega}^{z}{\wedge}{\omega}^{p}+{{\omega}^{p}}_{q}{\wedge}{\omega}^{q}+{{\omega}^{p}}_{z}{\wedge}{\omega}^{z}=0
\label{51}
\end{equation}
\begin{equation}
{\omega}^{z}{\wedge}{\omega}^{q}+{{\omega}^{q}}_{p}{\wedge}{\omega}^{p}+{{\omega}^{q}}_{z}{\wedge}{\omega}^{z}=0
\label{52}
\end{equation}
\begin{equation}
{{\omega}^{z}}_{q}{\wedge}{\omega}^{q}+{{\omega}^{z}}_{p}{\wedge}{\omega}^{p}=0
\label{53}
\end{equation}
From these relations one obtains the following Cartan connection one
forms
\begin{equation}
{{\omega}^{p}}_{z}=-{\alpha}{\omega}^{q}\label{54}
\end{equation}
\begin{equation}
{{\omega}^{q}}_{z}=-{\alpha}{\omega}^{p} \label{55}
\end{equation}
and
\begin{equation}
{{\omega}^{p}}_{q}=(1+{\alpha}){\omega}^{z} \label{56}
\end{equation}
where ${\alpha}$ is constant. Substitution of these connection form
components into the second Cartan equation
\begin{equation}
{R^{i}}_{j}={R^{i}}_{jkl}{\omega}^{k}{\wedge}{\omega}^{l}=d{{\omega}^{i}}_{j}+{{\omega}^{i}}_{l}{\wedge}{{\omega}^{l}}_{j}
\label{57}
\end{equation}
where ${R^{i}}_{j}$ is the Riemann curvature 2-form. After
straightforward algebra one obtains the following components of
Riemann curvature for the C-flows dynamo
\begin{equation}
{R^{q}}_{zzq}= -{\alpha}+{\alpha}^{2}\label{58}
\end{equation}
which is constant and negative if $0<{\alpha}<1$ and
\begin{equation}
{R^{q}}_{zzp}= -{\alpha}<0\label{59}
\end{equation}
while other components are easily computed. For Asonov geodesic
dynamo flows the important issue is to compute the Gaussian
curvature which is given by
\begin{equation}
d{{\omega}^{q}}_{z}={\alpha}{\omega}^{p}{\wedge}{\omega}^{z}
\label{60}
\end{equation}
\begin{equation}
d{{\omega}^{p}}_{z}=-{\alpha}{\omega}^{q}{\wedge}{\omega}^{z}
\label{61}
\end{equation}
 so the first scalar curvature is
$K_{1}={\alpha}$ , which by analogy yields $K_{2}=-{\alpha}$. Since
the Gaussian curvature is the product of $K_{1}$ and $K_{2}$ one
obtains $K_{G}=-{\alpha}^{2}<0$ and one certainly has a negative
curvature as necessarily for compact Riemannian Anosov manifolds.
\section{Conclusions}
 In conclusion, we obtain a class of C-flow dynamos in three-dimensional Riemannian manifold
 which generalizes the Arnold's Riemann metric. This solution of magnetic dynamo presents no pathologies in the
 and it is easy to show the magnetic flux and energy definitely grow which indicates that
 this solution can be considered a more realistic dynamo. More
 modern approaches to the dynamo problem , namely chaotic dynamos
 have been recently addressed by Reyl et al \cite{15} in the
 realm of plasma physics by investigating the quasi-two dimensional
 fast kinematic dynamo instabilities of chaotic fluid flow. This
 investigation, however is numerically and not analytically as was
 performed in this paper. Analytical solutions are still important
 to even guide us in building more general numerical solutions.
 Though our solution is more of a mathematical nature in certain
 sense without fold and reconnection \cite{16}, more realistic
 dynamo maps have considered recently \cite{17} and one could this
 path to build more general realistic dynamos.
\section{Acknowledgements} I
would like to thank CNPq (Brasil) and UERJ for financial supports.

\end{document}